\documentstyle[preprint,aps,epsf]{revtex}                  
\tightenlines
\begin{document}
\pagestyle{empty}                                      
\preprint{
\font\fortssbx=cmssbx10 scaled \magstep2
\hbox to \hsize{
\hfill$\raise .5cm\vtop{              
                \hbox{NTUTH-97-10}\hbox{NCTU-TH-97-04}}$}
}
\draft
\vfill
\title{Flavor Changing Neutral Higgs Couplings and 
Top-Charm Production at Next Linear Collider}

\vfill
\author{Wei-Shu Hou$^a$, Guey-Lin Lin$^b$ and Chien-Yi Ma$^{c}$ }
\address{
$^a$\rm Department of Physics, National Taiwan University,
Taipei, Taiwan, R.O.C.
}
\address{
$^b$\rm Institute of Physics, National Chiao Tung University,
Hsinchu, Taiwan, R.O.C.
}
\address{ $^c$\rm Department of  Electrophysics, 
National Chiao-Tung University, Hsinchu, 
Taiwan, R.O.C.}

%
%
\vfill
\maketitle
\begin{abstract}
We explore the possibility of detecting flavor changing neutral 
Higgs couplings at the Next Linear Collider (NLC) 
through $e^+e^-\to \nu_e\bar{\nu_e} t\bar{c}$.
In the framework of a general two-Higgs doublet model, 
we perform a complete calculation and find that
 $\sigma \left(e^+e^-\to \nu_e\nu_e t\bar{c},\; \nu_e\bar{\nu_e} {\bar t}c\right)$ 
could reach $\sim 9$ fb for $\sqrt{s}=2 \ {\rm TeV}$. 
This amounts to an annual production of 500 $t\bar{c}$ plus $\bar{t}c$ pairs 
at the NLC with an integrated luminosity of 50 fb$^{-1}$. The
dependence of $tc$-production rate on the neutral scalar mixing angle  
is mild except when $\sin^2\alpha \to 0 \ {\rm or} \ 1$.
The $\nu\nu W^+W^-$ background should be manageable 
after $b$-tagging, while $\nu\nu t\bar t$ background should not be a problem 
when the signal event rate is still interesting.
The process, together with 
$e^+e^-\to \nu_e\bar{\nu_e} W^+W^-,\; \nu_e\bar{\nu_e} ZZ$
studies, offer the chance of measuring the $t$-$c$-Higgs coupling.

\end{abstract}
%
%
\pacs{PACS numbers:
14.80.Cp, 14.65.Ha, 12.15.Ff, 13.90.+i  }
%
%
\pagestyle{plain}

\section{Introduction}
 
The mechanism for symmetry breaking and 
the fermion mass and mixing hierarchy pattern 
are the two remaining mysteries in the electroweak theory.
The construction of high energy colliders 
such as the LHC and NLC are in fact aimed at resolving such mysteries. 
In this regard, the physical processes that should be studied thoroughly 
at such  machines are those involving the top quark, 
whose properties have yet to be studied carefully,
as well as the yet to be discovered Higgs boson(s).

It was suggested some time ago that\cite{CS}, in multi-Higgs doublet models, 
the ``natural" flavor conservation condition \cite{GW} is not mandatory for 
the suppression of flavor changing neutral current (FCNC) processes. 
Rather, Nature has provided its own cure:
the existing hierarchical patterns in quark masses and mixing angles
may imply a pattern for flavor changing neutral Higgs couplings (FCNH) 
that is consistent with low energy data \cite{CS}.   
An interesting consequence of this framework is the possibility of
sizable $t$-$c$-neutral Higgs couplings which would have
notable impact on top quark and Higgs physics \cite{Hou,HW}.  
To probe such couplings at colliders, several processes \cite{ARS,HL,HLMY}
have been proposed which can be studied at the NLC or LHC.
At the NLC, one may look for $t\bar c$ pair production
via $e^+e^-\to Z^*\to t\bar{c},\; \bar{t}c$\cite{ARS} 
(where the $Z$-$t$-$c$ coupling is loop-induced),
or like-sign top pair production via 
$e^+e^-\to h^0A^0\to tt\bar{c}\bar{c},\; \bar{t}\bar{t}cc$ \cite{HL}.
At the LHC, such flavor non-diagonal couplings
can be probed through the parton subprocess 
$cg\to tA^0\to tt\bar{c}$ \cite{HLMY}, which involves the FCNH coupling
directly in the production process. 

Recently, Bar-Shalom {\it et  al.} pointed out \cite{shalom} that 
FCNH couplings may be probed at the NLC via the $WW$ fusion process
$e^+e^-\to \nu_e\bar{\nu}_e t\bar{c},\; \nu_e\bar{\nu}_e\bar t c$, as shown in Fig. 1. 
With $\sqrt{s}=2 \ {\rm TeV}$, and the masses of 
neutral Higgs bosons being 250 GeV and 1 TeV respectively, they found 
$\sigma_{\nu\nu tc} \equiv \sigma 
\left(e^+e^-\to \nu_e\bar{\nu}_e t\bar{c}\right)
+\sigma \left(e^+e^-\to \nu_e\bar{\nu}_e \bar{t}c\right) \approx$ 5 fb. 
Assuming an integrated luminosity of 50 fb$^{-1}$ at the NLC, 
this implies an annual production of 125 $t\bar{c}$
and an equal number of $\bar{t}c$ pairs. 
The process has a much larger $tc$ production rate 
than $e^+e^-\to Z^*\to t\bar{c}$, 
and does not suffer from $s$-channel suppression
as $e^+e^-\to h^0A^0\to tt\bar{c}\bar{c}$. 
In view of this, we would like to follow up on this work.
We shall perform a full calculation and compare with 
the effective $W$ approximation used in Ref. [8],
explore different scenarios for neutral Higgs masses, 
and clarify parameter dependence of the $tc$ production cross section.       

The paper is organized as follows: In Section II,
we briefly review the two-Higgs doublet model 
with FCNH couplings and present the result of 
a full calculation of $\sigma_{\nu\nu tc}$ using helicity methods. 
We then point out that $\sigma_{\nu\nu tc}$ is
largest when both neutral scalars 
have mass of order the weak scale.
This becomes the focus of our discussion throughout the paper.
In Section III, we demonstrate the utility of the narrow width approximation.
In Section IV we show that $\sigma_{\nu\nu tc}$ is not sensitive to 
the mixing angle of neutral scalars, 
and remains at the fb level for $\sqrt{s} \ge 1$ TeV.
Some discussion of signal vs. background is given.
After concluding in Section V, we leave some technical details 
in Appendices A and B.

\section{Full Calculation}

The calculation of $\sigma_{\nu\nu tc}$ is based on the Lagrangian of 
a general two-Higgs doublet model with flavor changing
neutral Higgs (FCNH) couplings   
\begin{eqnarray}
{\cal L} = && (D_{\mu}\Phi_1)^{\dagger}(D^{\mu}\Phi_1)
                     +(D_{\mu}\Phi_2)^{\dagger}(D^{\mu}\Phi_2) - V(\Phi_1,\Phi_2)\nonumber \\
&\ - \ & \left( \bar u_L {M}^u u_R + \bar d_L{M}^d d_R \right)\,
            \sqrt{2}\, {\mbox{Re}\, \phi_1^0\over v} \nonumber \\
    &\ + \ & \left(\bar u_{L} \xi^{\left(u\right)} u_{R}
+ \bar d_{L} \xi^{\left(d\right)} d_{R}\right)\,
\sqrt{2}\, \mbox{Re}\, \phi_2^0 \nonumber \\
      &\ + \ & \left(-\bar u_{L} \xi^{\left(u\right)} u_{R}
               + \bar d_{L} \xi^{\left(d\right)} d_{R}\right)\,
                 i\,\sqrt{2}\, \mbox{Im}\, \phi_2^0 \nonumber \\
   &\ - \ & \bar d_{L} V^\dagger \xi^{\left(u\right)} u_{R}\,  \sqrt{2}\, \phi_2^-
       + \bar u_{L} V\,       \xi^{\left(d\right)} d_{R}\, \sqrt{2}\, \phi_2^+
                \ + \mbox{H.c.},
\end{eqnarray}
where $u_{L,(R)}$ and $d_{L,(R)}$ are flavor 
multiplets of up-type and down-type quarks respectively, and 
${M}^{u,d}$ are their diagonalized mass matrices. 
Note that we have relegated all the FCNH couplings 
to the second doublet as a result of rotating to the specific basis 
$\left\langle \phi_2^0 \right\rangle =0$ and
$\left\langle \phi_1^0 \right\rangle =v/ \sqrt{2}$\cite{LS}.   
This is because there is no discrete symmetry \cite{GW} 
as in usual two Higgs doublet models \cite{Hguide} 
to distinguish between $\Phi_1$ and $\Phi_2$,
so the familiar $\tan\beta \equiv v_1/v_2$ parameter is not physical.
Assuming CP invariance in the Higgs sector, the scalar fields 
$\sqrt{2}\,\mbox{Im}\, \phi_2^0$ and $\phi_2^\pm$ are identified as 
the physical pseudoscalar boson $A^0$ and charged scalar $H^{\pm}$.
The CP even neutral scalars $\sqrt{2}\, \mbox{Re}\,\phi_1^0$ 
and $\sqrt{2}\, \mbox{Re}\,\phi_2^0$ can still mix through the Higgs potential 
$V(\Phi_1,\Phi_2)$ into the physical states $H^0$ and $h^0$. 
In the limit that the mixing angle $\sin\alpha \to 0$,
$H^0 \leadsto \sqrt{2}\, \mbox{Re}\, \phi_1^0$
becomes the ``standard" Higgs boson with diagonal couplings,
while $h^0 \leadsto \sqrt{2}\, \mbox{Re}\, \phi_2^0$
has FCNH couplings characterized by the 
non-diagonal matrix $\xi^{u,d}_{ij}$.

In our calculation as well as in Ref. [8], the simple Cheng-Sher
ansatz\cite{CS} is adopted:
\begin{equation}
\xi^{u,d}_{ij} =f_{ij}\, {\sqrt{m_i m_j} \over v},
\end{equation}
where $f_{ij}$'s are constants of order unity. 
The coupling $\xi^{u}_{tc}$ is expected to be the largest and has the
most prominent signature to be searched for in collider experiments. 
From Eq. (1), we can now single out the relevant couplings for computing
the process $e^+e^-\to \nu_e\bar{\nu}_e t\bar{c},\; 
\nu_e\bar{\nu}_e {\bar t} c$ given in Fig. 1. 
Since we wish to compare with Ref. [8], we take $f_{tc} \simeq \sqrt{2}$. The resulting 
$t$-$c$-{\rm Higgs} and Higgs-$W$-$W$ couplings are 
\begin{equation}
{\cal L}_{int.}={g\sqrt{m_tm_c}\over \sqrt{2}m_W}\, 
(\sin\alpha\, H + \cos\alpha\, h)\; \bar{t}c + \mbox{h.c.} 
                         \ + \ g\, m_W\, (\cos\alpha\, H - \sin\alpha\, h)\;  W^{\mu}W_{\mu}.
\end{equation}
The Higgs-$Z$-$Z$ couplings can be easily incorporated,
and the cross sections for $e^+e^-\to e^+e^- t\bar{c},\; 
e^+e^- {\bar t} c$ via $ZZ$ fusion are simply related to that of
 $e^+e^-\to \nu_e\bar{\nu}_e t\bar{c},\; 
\nu_e\bar{\nu}_e {\bar t} c$ \cite{shalom}.

\subsection{Helicity Amplitude Calculation}

A full calculation of $\sigma_{\nu\nu tc}$ is rather involved 
as the process considered is a $2\to 4$ scattering. 
An efficient way of doing it is by employing the helicity method \cite{ONETOP},
which facilitates the numerical manipulations of Feynman amplitudes. 

The amplitude for $e^+(p_1)e^-(p_2)\to  \bar{\nu}_e(p_3) \nu_e(p_4) 
t(p_t)\bar{c}(p_c)$ reads
\begin{eqnarray}
i{\cal M} & = &
   F\; \left[\bar{v}(p_1,\lambda_1)\gamma_{\mu}
                  P_{-} v(p_3,\lambda_3)\right]\; 
          \left[\bar{u}(p_4,\lambda_4)\gamma^{\mu}
                  P_{-} u(p_2,\lambda_2)\right]\;
          \left[\bar{u}(p_t,\lambda_t) v(p_c,\lambda_c)\right]\nonumber \\
& \times & \left[ {i\over q^2-m_H^2+im_H\Gamma_H}
-{i\over q^2-m_h^2+im_h\Gamma_h} \right]
{-i\over (p_1-p_3)^2-m_W^2}\; {-i\over (p_2-p_4)^2-m_W^2},
\end{eqnarray} 
where $q$ is the momentum of the intermediate 
Higgs boson, $P_{\pm}\equiv (1\pm \gamma_5)/2$, and 
\begin{equation}
F=\cos\alpha\sin\alpha\; \left({ig\over \sqrt{2}}\right)^2 (ig m_W)
\left({ig \sqrt{m_tm_c}\over \sqrt{2}m_W}\right),
\end{equation}
is a collection of coupling coefficients.
Note that, except for the relative sign and differences in mass and width,
the $h$ and $H$ contributions are basically the same.
All fermion masses are set to zero except for the top quark, 
and the $m_c$ dependence is kept only in the coupling of Eq. (2).
The helicities of leptons are therefore completely fixed by their 
left-handed vector couplings to $W$ bosons, i.e.
$\lambda_1=\lambda_3=+$ and $\lambda_2=\lambda_4 =-$. 
However, there are four combinations involving the helicities of 
top and charm quarks. 

Let $A(\lambda_1,\lambda_3)\equiv \bar{v}(p_1,\lambda_1)
\gamma_{\mu} P_{-} v(p_3,\lambda_3)$ and
$B(\lambda_2,\lambda_4)\equiv \bar{u}(p_4,\lambda_4)
\gamma^{\mu} P_{-} u(p_2,\lambda_2)$.
One finds (see Appendix A for details)
\begin{eqnarray}
A(++)&=& \sqrt{2E_1}\sqrt{2E_3}\left\langle \hat{p}_3+ \vert \gamma_{\mu -} \vert \hat{p}_1+ \right\rangle ,\nonumber \\
B(--)&=& \sqrt{2E_2}\sqrt{2E_4}\left\langle \hat{p}_4- \vert \gamma_+^{\mu} \vert \hat{p}_2- \right\rangle,
\end{eqnarray}
where
$\omega_{\pm t}= \sqrt{E_t\pm \vert \vec{p}_t \vert}$, 
$\gamma^{\mu}_{\pm}$ are $2\times 2$ matrices 
defined by $\gamma^{\mu}_{\pm}=\left(1,\mp \vec{\sigma}\right)$, 
and $\vert \hat{p}\pm \rangle$ denote the two-component eigenvectors of 
the helicity operator $\vec{p}\cdot \vec{\sigma}/ \vert \vec{p} \vert$,
that is
\begin{equation}
\vert \hat{p}+\rangle=\left (\matrix{
\cos{\theta\over 2}\cr
e^{i\phi}\sin{\theta\over 2}\cr
}\right ), \ \ 
\vert \hat{p}-\rangle=\left (\matrix{
-e^{-i\phi}\sin{\theta\over 2}\cr
\cos{\theta\over 2}\cr
}\right ),
\end{equation}
where $\theta$ and $\phi$ are angles for $\vec{p}$.
For the top-charm scalar density, define
$C(\lambda_t,\lambda_c)\equiv \bar{u}(p_t,\lambda_t) v(p_c,\lambda_c)$,
one gets four combinations (see Appendix A for details)
\begin{eqnarray}
C(\mp\mp)&=&-\sqrt{2E_c}\; \omega_{+t}\; 
                    \left\langle \hat{p}_t\mp\vert \hat{p}_c\pm \right\rangle , \nonumber \\
C(\mp\pm)&=&-\sqrt{2E_c}\; \omega_{-t}\; 
                   \left\langle \hat{p}_t\mp\vert \hat{p}_c\mp \right\rangle .
\end{eqnarray}
Since $A(++)$ and
$B(--)$
are already fixed, there are four helicity amplitudes
$i{\cal M}(\lambda_t,\lambda_c)\propto C(\lambda_t,\lambda_c)$. 
With all four helicity amplitudes constructed,
the subsequent numerical calculations can be done in a
straightforward manner by utilizing the program ONETOP \cite{ONETOP}.

\subsection{Comparison with Bar-Shalom {\it et al.}}

To compare with Ref. [8], we compute $\sigma_{\nu\nu tc}$ 
for $m_H= 1 \ {\rm TeV}$ and $\sin^2\alpha={1/2}$. 
The cross section $\sigma_{\nu\nu tc}$ as a function of $m_h$ 
for $\sqrt{s}=$ 0.5, 1, 1.5 and 2 TeV are shown in Fig. 2. 
It peaks notably at $m_h\approx 250 \ {\rm GeV}$ and 
decreases monotonously as $m_h$ increases from 250 GeV. 
In accordance with the difference in the propagators given in Eq. (4),
it vanishes in the degenerate limit $m_h=m_H=1 \ {\rm TeV}$.
This is a special case for the choice of $\sin^2\alpha=1/2$ (i.e. $\alpha={\pi/ 4}$),
because the Higgs properties are identical in the degeneracy limit, 
so the amplitudes arising from each Higgs would then cancel completely.  
For $\sqrt{s}=2 \ {\rm TeV}$, the maximal value of
$\sigma_{\nu\nu tc}$ is around 4.5 fb, which is smaller than 5.2 fb 
obtained in Ref. [8] which uses the effective $W$ approximation. 
Such an overestimation by 
the effective $W$ approximation is a typical phenomenon 
in collider physics\cite{Dawson}. 

The prominent peaks in Fig. 2 suggest that the cross section 
arising from $h$ alone would be the largest at $m_h\sim 250 \ {\rm GeV}$. 
Similar behavior should then be expected for the contribution from $H$. 
We therefore expect the total cross section resulting 
from $H$ and $h$ to be the largest 
{\it if both $m_H$ and $m_h$ are of order the weak scale}.
This precisely fits the arguments given in Refs. [6,7] which 
emphasizes the mass range
\begin{equation}
200 \ {\rm GeV}< m_h, \ m_H < 2m_t\approx 350 \ {\rm GeV}.
\end{equation}  
The lower bound is to allow the $t\bar c$ threshold to turn on. 
The upper bound of $2m_t\approx 350 \ {\rm GeV}$ 
was imposed originally for the pseudoscalar $A^0$.
For $h$ and $H$, as can be seen from Fig. 2,
the cross section is still sizable up to $m_{h,\; H} \cong 500$ GeV
for $\sqrt{s} > 1$ TeV.
This is because $\Gamma(h,\; H\to VV)\gg
\Gamma(h,\; H\to f'\bar{f})$ for $m_{h,\; H} \sim 350 \ {\rm GeV}$, 
and the opening of $t\bar{t}$ mode does not increase
substantially the total width of $h$ or $H$.   
However, for the range of Eq. (9), the $t\bar{t}$ 
background to the $t\bar{c}$ or $\bar{t}c$ modes would be suppressed.

To show that $\sigma_{\nu\nu tc}$ is indeed 
more significant in the the range of Eq. (9), 
we show in Fig. 3 the cross section $\sigma_{\nu\nu tc}$  
as a function of $m_h$ for $m_H=300 \ {\rm GeV}$ and $\sin^2\alpha={1/2}$. 
The cross section drops to zero at the degenerate limit 
$m_h=m_H=300 \ {\rm GeV}$ in a much more dramatic way.
However, such a severe cancellation does not generally occur 
since there is no reason for $m_h$ and $m_H$ to be degenerate,
and the cancellation is anyway incomplete for other values of $\sin\alpha$.
The cancellation effect is negligible if the mass difference 
$\Delta M = \vert m_H - m_h\vert$ is a  few times 
the widths of both Higgs bosons (see Appendix B). 
Slightly away from the degeneracy limit, 
the cross section rises to its peak value $\cong 8.0$ fb 
at $m_h\approx 250 \ {\rm GeV}$ for $\sqrt{s}=2 \ {\rm TeV}$,
which is almost twice as large as the case with $m_H=1 \ {\rm TeV}$. 
As $m_h$ increases to 1 TeV, $\sigma_{\nu\nu tc}$ drops 
to about $3.6$ fb, which is mostly from $H$. 
For a lighter $h$, i.e. $m_h < 250 \ {\rm GeV}$, 
the cross section also drops. 
This once again illustrates the fact that $\sigma_{\nu\nu tc}$ 
receives the largest individual contributions from $h$ and $H$ 
respectively at $m_h, \  m_H\approx 250 {\rm\ GeV}$.

\section{The narrow width approximation}

It is important to note that the widths of neutral Higgs in
the mass range of Eq. (9), even up to $\sim 500$ GeV,
are still quite small compared to their masses.
The Standard Model (SM) Higgs boson ${H_{\rm SM}}$ provides 
an upper bound to $H$ and $h$ widths, for example
$\Gamma_{H_{\rm SM}}\approx 15 \ {\rm GeV}$ 
for $m_{H_{\rm SM}}=350 \ {\rm GeV}$ \cite{Hguide}. 
Since the widths of both Higgs are narrow in the mass range of interest, 
it is convenient to compute $\sigma_{\nu\nu tc}$ in 
the narrow width approximation. 
We may approximate $\sigma_{\nu\nu tc}$ by the cross section of
Higgs production $\sigma (e^+e^-\to \bar{\nu_e}\nu_e h(H))$ 
multiplied by the branching ratio of the flavor changing decay
$h(H)\to t\bar{c},\; \bar{t}c$. 
This approach is much simpler than the previous full calculation or
even the effective $W$ approximation.
One can then determine the Higgs mass and $\sin^2\alpha$
dependences of $\sigma_{\nu\nu tc}$ with ease.

\subsection{$WW\to h,\; H$ Production}

Compared to the previous calculation of $\sigma(e^+ e^- \to \nu\bar \nu t\bar c)$,
it is considerably simpler to compute the cross section 
$\sigma(e^+e^- \to \bar{\nu}_e \nu_e h(H))$. 
It is identical to that of SM Higgs production $\sigma(e^+e^-\to \bar{\nu_e}\nu_e 
{H_{\rm SM}})\equiv \sigma_{\nu\nu {H_{\rm SM}}}$ \cite{Hguide}, 
except for the additional factors of $\cos^2\alpha$ or $\sin^2\alpha$. 
The amplitude for $e^+e^-\to {\bar \nu_e}\nu_e {H_{\rm SM}}$ is
\begin{eqnarray}
{\cal M}(e^+(p_1)e^-(p_2)&\to& {\bar \nu_e}(q_1)\nu_e(q_2) {H_{\rm SM}}(k))= {ig^3m_W
\over 8}\left({1\over 2p_1\cdot q_1+m_W^2}\right)\left({1\over 
2p_2\cdot q_2 +m_W^2}\right)\nonumber \\
&\times &\left[\bar{v}(p_1,s_1)\gamma^{\mu} (1-\gamma_5)v(q_1,s_2)\right]\;
                  \left[\bar{u}(q_2,s_4)\gamma_{\mu} (1-\gamma_5)u(p_2,s_3)\right].
\end{eqnarray}
Averaging over the initial and summing over the final state spins give
\begin{equation}
{1\over 4}\sum_{pol.}{\vert {\cal M}\vert}^2
=g^6m_W^2\left({1\over 2p_1\cdot q_1+m_W^2}\right)^2\left({1\over 
2p_2\cdot q_2 +m_W^2}\right)^2(p_2\cdot q_1)
(p_1\cdot q_2),
\end{equation}
where we have neglected fermion masses. 
The final state phase space integration is done by VEGAS \cite{vegas}. 
For $\sqrt{s}=2 \ {\rm TeV}$ and $m_{H_{\rm SM}}=250 \ {\rm GeV}$,
we find $\sigma(e^+e^-\to \bar{\nu_e}\nu_e {H_{\rm SM}})\approx 264 \ {\rm fb}$. 
The cross section for other values of $m_H$ and $\sqrt{s}$ 
can be read off  from Fig. 4.

\subsection{$h,\; H \to t\bar c$ Decay}

To compute the branching ratio BR$(h,\; H\to tc)$,    
we note that the dominant decay modes 
for $m_{h,\; H}<2m_t$ are $h,\; H\to W^+W^-, ZZ, b\bar{b}$ \cite{Hguide} 
and $\ t\bar{c}$, $\bar{t} c$\cite{Hou},
where the latter are specific to the current model.
The width of each decay mode is well known: 
\begin{eqnarray}
\Gamma (H\to W^+W^-)&=&{g^2m_H^3\over 64\pi m_W^2}     
\cos^2\alpha\,\sqrt{1-4x_W^2}\; (1-4x_W^2+12 x_W^4),
\nonumber\\
\Gamma (H\to ZZ)&=&{g^2m_H^3\over 128\pi m_W^2}     
\cos^2\alpha\,\sqrt{1-4x_Z^2}\; (1-4x_Z^2+12x_Z^4),
\nonumber\\
\Gamma (H\to b\bar{b})
&\simeq& {3g^2m_H\over 32\pi m_W^2}\; m_b^2\; 
(1-4x_b^2)^{3/2}, \nonumber \\
\Gamma (H\to t\bar{c}+\bar{t}c)
&=& \left({f_{tc}\over \sqrt{2}}\right)^2\times {3g^2m_H\over 8\pi m_W^2}\; m_tm_c \; \sin^2\alpha\,
(1-x_+^2)^{3/2}\; (1-x_-^2)^{1/2},
\end{eqnarray}   
with $x_i={m_i/ m_H}$ and $x_\pm={(m_t\pm m_c)/ m_H}$. 
For $\Gamma (h\to W^+W^-, ZZ)$ , etc.
one simply changes $\sin^2\alpha\to \cos^2\alpha$.
Note that we have assumed SM couplings for $b\bar b$,
although it should depend on more parameters
 (this is another reason for the mass range of Eq. (9) 
so we avoid uncertainties in $H(h)$-$t$-$\bar t$ coupling).
However, the $b\bar b$ mode is unimportant for our purpose.

For a generic mixing angle
$\alpha$, vector boson decay modes  
dominate over the fermionic ones since the
former is proportional to $m_{h,\; H}^3$ while the
latter only depends on $m_{h,\; H}$ linearly. 
One can clearly see in Fig. 1 of Ref.  \cite{HL} this severe suppression 
of BR$(h,\; H\to tc)$ for generic $\alpha$ values \cite{NOTE}. 
However, for extreme values of $\alpha \to 0$ or $1$, 
the $WW$, $ZZ$ modes could be very suppressed, 
and either BR$(h\to tc)$ or BR$(H\to tc)$ become significant \cite{HL}.  

The threshold behavior of the $t\bar c$ mode and the 
dominance of $h,\; H\to WW,\ ZZ$ modes in general help us understand 
the peak in $\sigma_{\nu\nu tc}$ at $m_{h,\; H} \approx 250 \ {\rm GeV}$.
We show in Fig. 5  the mass dependence of BR$(h\to tc)$ for 
a few values of $\sin^2\alpha$ in the range
\begin{equation}
0.1 < \sin^2\alpha <0.9.
\end{equation}
BR$(H\to tc)$ can be simply obtained by making the change
$\sin^2\alpha \to \cos^2\alpha$.
We do not include extreme cases of $\sin^2\alpha \to 0 \ {\rm or} \ 1$ 
since $\sigma_{\nu\nu tc}\to 0$ in these limits.  
The shape of Fig. 5 can be understood as follows.
BR$(h,\; H\to tc)$ rises sharply right after the 
opening of the $tc$ production threshold. 
The growth of BR$(h,\; H\to tc)$ should however stop at 
certain Higgs mass, since $\Gamma(h,\; H\to VV)$ is in general dominant
and grows more rapidly as $m_{h,\; H}$ increases. 
The peak position $m_{h,\; H}\cong 260 \ {\rm GeV}$ for BR$(h,\; H\to tc)$,
which is the main reason behind the peaks seen in Figs. 2 and 3,
marks the point where the growth in $\Gamma(h,\; H\to tc)$ 
is overtaken by $\Gamma(h,\; H\to VV)$. 
It is interesting to note that 
BR$(h,\; H\to tc)$ always peaks at  $m_{h,\; H}\cong 260 \ {\rm GeV}$ 
independent of the $\sin^2\alpha$ we choose.
This is easily understood since, for generic $\alpha$, 
BR$(H\to tc)\equiv \Gamma(H\to tc)/ \Gamma_H 
\approx \Gamma(H\to tc)/ \Gamma(H\to VV)$, 
i.e. BR$(H\to tc)\approx \tan^2\alpha\; f(m_H)$ 
where $f(m_H)$ is largely $\alpha$-independent 
and peaks at $m_H\cong260 \ {\rm GeV}$.  
Similarly, we have BR$(h\to tc)\approx \cot^2\alpha\; f(m_h)$. 
Such a simple dependence on the mixing angle  $\alpha$ makes Fig. 5 very useful. 
For any $\sin^2\alpha$ in the range Eq. (13),
one can obtain the branching ratio BR$(h,\; H\to tc)$ 
for any Higgs mass by simply scaling via the relation 
BR$(h,\; H \to tc)\approx \cot^2\alpha\, f(m_h^2),\, \tan^2\alpha\, f(m_H^2)$.    

We note that the kink due to $t\bar t$ threshold becomes
more visible for small $\sin^2\alpha$ values.
This is because the $VV$ contribution to the Higgs width 
becomes suppressed and the relative weight of 
the $t\bar t$ contribution becomes more significant \cite{NOTE2}.
Such a kink is not apparent in Figs. 2 and 3 because 
the $\sin^2\alpha = 1/2$ case was used.

\subsection{Cross Section}

The SM Higgs width provides an upper bound to
$\Gamma_H$ and $\Gamma_h$.
We can therefore use the narrow width approximation
for $m_{H,\, h} < 500$ GeV.
The cross section $\sigma_{\nu\nu tc}$ can be written as
\begin{eqnarray}
\sigma_{\nu\nu tc}&\cong &
\sigma_{\nu\nu {H_{\rm SM}}}(m_{H_{\rm SM}}=m_H)
\times \cos^2\alpha\times \mbox{BR}(H\to tc)  \nonumber \\
&+& \sigma_{\nu\nu {H_{\rm SM}}}(m_{H_{\rm SM}}=m_h)
\times \sin^2\alpha\times \mbox{BR}(h\to tc)\ \ + \ \ {\rm interference \ terms},
\end{eqnarray}
where ${H_{\rm SM}}$ denotes SM Higgs.
Note that, with $\vert m_H^2-m_h^2\vert > (3-4)\times m_h\Gamma_h$, 
the interfernce term can be safely neglected (see Appendix B for details).

To locate the peak of $\sigma_{\nu\nu tc}$ for generic $\sin^2\alpha$, 
Eq. (14) can be rewritten as
\begin{equation}
\sigma_{\nu\nu tc}\cong\sigma_{\nu\nu {H_{\rm SM}}}
(m_H)\times \sin^2\alpha\times f(m_H) 
+ \sigma_{\nu\nu {H_{\rm SM}}}
(m_h)\times \cos^2\alpha\times f(m_h),
\end{equation}
where we have neglected the interference term by assuming 
a large enough splitting between $m_H$ and $m_h$. 
With $m_H$ and $\sin^2\alpha$ fixed as in the case of Figs. 2 and 3, 
$\sigma_{\nu\nu tc}$ only depends on 
$\sigma_{\nu\nu {H_{\rm SM}}} (m_h)\times f(m_h)$. 
Since $f(m_h)$ peaks at $m_h=260 \ {\rm GeV}$ and 
$\sigma_{\nu\nu {H_{\rm SM}}}$ is a monotonously decreasing function of $m_{H_{\rm SM}}$, 
the position of $m_h$ giving maximal $\sigma_{\nu\nu tc}$ should be 
shifted downward from 260 GeV. 
This is exactly the case as seen in Figs. 2 and 3 where such effect 
are most significant for $\sqrt{s}=0.5 \ {\rm TeV}$ since
$\sigma_{\nu\nu {H_{\rm SM}}}$
drops most steeply for increasing $m_{H_{\rm SM}}$ for this case. 
For $\sqrt{s}=2 \ {\rm TeV}$, this shift becomes much smaller 
as $\sigma_{\nu\nu {H_{\rm SM}}}$ is relatively flat.

\section{Discussion}

To illustrate our arguments so far, let us explore the ``maximal" and ``minimal"
$\sigma_{\nu\nu tc}$ cross sections in the mass range of Eq. (9),
and $\sin\alpha$ and $\sqrt{s}$ dependences.
We shall also make some general discussions about signal vs. background
and compare with other processes.

\subsection{Range of Cross Sections}

For ``maximal" $\sigma_{\nu\nu tc}$, take , for example, 
$m_H=250 \ {\rm GeV}$ and $m_h=240 \ {\rm GeV}$ so
$\vert m_H^2-m_h^2\vert 
{\ \lower-1.2pt\vbox{\hbox{\rlap{$>$}\lower5pt\vbox{\hbox{$\sim$}}}}\ }
4\times m_h\Gamma_h$,
and the interference term in $\sigma_{\nu\nu tc}$ can be safely neglected. 
Since the masses are approximately equal, one can rewrite Eq. (14) as
\begin{equation}
\sigma_{\nu\nu tc}\approx
\sigma_{\nu\nu {H_{\rm SM}}}
\times \left( \cos^2\alpha\; \mbox{BR}(H\to tc)
+\sin^2\alpha\; \mbox{BR}(h\to tc)\right),
\end{equation}
where the mass of ${H_{\rm SM}}$ can be taken as either that of $H$ or $h$.
Note that the combination 
$ \cos^2\alpha\, \mbox{BR}(H\to tc)+\sin^2\alpha\, \mbox{BR}(h\to tc)$ 
determines the $\sin^2\alpha$ dependence of $\sigma_{\nu\nu tc}$,
which is plotted in Fig. 6. 
It is interesting to see that both $\cos^2\alpha\; \mbox{BR}(H\to tc)$ 
and $\sin^2\alpha\; \mbox{BR}(h\to tc)$ are sensitive to $\sin^2\alpha$ 
but their sum is not.
This is in large part because we chose almost equal $m_H$ and $m_h$,
and reflects the mutually compensating nature between the two contributions.
The effective fraction 
$\cos^2\alpha\, \mbox{BR}(H\to tc)+\sin^2\alpha\, \mbox{BR}(h\to tc)$
of the (``SM") Higgs production cross section stays between $2-3\%$ 
for almost the entire range of $\sin^2\alpha$ of Eq. (13),
but becomes extremely suppressed for $\sin^2\alpha$ outside this range..  
 
For $\sqrt{s}=2 \ {\rm TeV}$, $\sin^2\alpha=1/2$, and 
$m_H,\, m_h=250,\, 240 \ {\rm GeV}$, 
 from Figs. 4 and 6 we find
\begin{equation}
\sigma_{\nu\nu tc}\approx 270 \ {\rm fb}\times 3.2\%
=8.6 \ {\rm fb}.
\end{equation}  
This is in good agreement with the maximal cross section obtained earlier
from the full calculation, 
and illustrates the effectiveness of the narrow width approximation.
The $\sin^2\alpha$ dependence is very mild. 
For example, at $\sin^2\alpha=0.1 \ {\rm or} \ 0.9$,
$\sigma_{\nu\nu tc}=6.8$ fb for $\sqrt{s}=2 \
{\rm TeV}$, which is still comparable to the maximal cross section.  
The $\sin^2\alpha$ dependence for individual $h$ or $H$ contributions
is much more significant.

To explore the ``minimal" cross sections within the range of Eq. (9),
we note from Figs. 2 and 3 that the contribution of 
$h,\, H$ to $\sigma_{\nu\nu tc}$ is roughly equal for $m_{h,\; H}=200 \ {\rm GeV}$
and $m_{h,\; H}=350 \ {\rm GeV}$. 
We therefore present the results for $m_H=350 \ {\rm GeV}$ 
and $m_h=200 \ {\rm GeV}$, which gives roughly 
the smallest $\sigma_{\nu\nu tc}$ for the mass range of interest. 
We plot in Fig. 7  $\sigma_{\nu\nu tc}$ for this set of Higgs masses 
as a fuction of $\sin^2\alpha$ for $\sqrt{s}=0.5, \ 1, \ 1.5, \ 2.0 \ {\rm TeV}$. 
It is seen that $\sin^2\alpha$ dependence remains mild.
What is remarkable is that, 
{\it for almost all values of $\sin^2\alpha$, 
$\sigma_{\nu\nu tc}$ is at fb level  or higher for $\sqrt{s}\geq 1 \ {\rm TeV}$}. 
This promising result for $\sigma_{\nu\nu tc}$ holds only 
in the mass range given by Eq. (9), although the range can be extended up to 
$m_{h,\; H} \sim 400$--$450$ GeV or so. 

In both Figs. 6 and 7 we have illustrated with cases where 
the $h$ and $H$ peak (in $\sin^2\alpha$) contributions are comparable,
hence their $\sin^2\alpha$ dependences are mutually compensating. 
For more general choices of $m_h$ and $m_H$ values,
some $\sin^2\alpha$ dependence would remain for $\sigma_{\nu\nu tc}$,
which is reflected in and easily scaled from the individual $h$ or $H$ contributions.

\subsection{Signal {\it vs.} Background}

Turning to the experimental signal at the NLC, 
one needs to consider the final states from top decay,
$t\to b\ell^+\nu,\; bj_1 j_2$, hence the signal modes are
\begin{equation}
e^+e^-\longrightarrow \nu_e\bar{\nu}_e t\bar{c}
\longrightarrow \nu_e\bar{\nu}_e \nu_\ell\; \ell^+ b\bar c,\; 
                             \nu_e\bar{\nu}_e\, b\bar{c}j_1j_2,
\end{equation}
and similarly for $\nu_e\bar{\nu}_e \bar{t}c$.
Since typical cross sections are a few fb in the mass range of Eq. (9),
with an integrated luminosity of 50 fb$^{-1}$, 
we expect of order 100 or so (no more than 300) $\nu\nu bcj_1j_2$ events,
and 1/6 of this in each $3\nu + \ell bc$ channels, 
where $\ell = e^\pm,\; \mu^\pm,\; \tau^\pm$.
Although the event rates are significant, we find that 
the latter is not very promising once backgrounds are taken into account.

What are the potential backgrounds? 
Since the $\nu_e\bar{\nu}_e$ pair should carry away
missing transverse energy ${E \hskip -0.45cm \not \hskip 0.35cm }_T  \sim m_W$,
$WW$ fusion events should be relatively distinct at the NLC.
For the mass range $m_{h,\; H} < 2m_t$ of Eq. (9), background from 
$e^+e^-\to \nu_e\bar{\nu}_e h,\, \nu_e\bar{\nu}_e H 
\to \nu_e\bar{\nu}_e t\bar{t}$ is absent. 
The major background to be considered is therefore    
$e^+e^-\to \nu_e \bar{\nu}_eW^+W^-$ since it is more abundantly produced
via ${h,\; H} \to W^+W^-$. 
From Figs. 4--7 one sees that  
$\sigma(e^+e^-\to \nu_e\bar{\nu}_e W^+W^-)$ is typically $20$
to  $30$ times larger than the signal.
But the important point to notice is that $W$ decays 
do not contain $b$ quarks ($< 10^{-3}$ in BR).
The chief tool to suppress the $WW$ background is therefore $b$-tagging, 
expected to be very efficient at Linear Colliders \cite{btag}.
However, since 1/3 of $W$ decays contain charm quarks,
fake rate of $b$-tagging might be an issue.
In particular, the $3\nu + \ell + bc$ mode would not be easy to distinguish from
$3\nu + \ell + cs$ fakes when the signal event rate is so low.
In contrast, the $\nu\nu\; b c j_1 j_2$ mode has a second handle: 
kinematics and full reconstruction.
With one $b$-tagged jet, two of the three remaining jets should reconstruct to 
$m_W$ \cite{jet}, and together with the $b$-jet reconstruct to a top quark.
After such reconstruction, the signal events should show a mass peak
over the $WW$ background.
Note that the $WW$ ``background" is itself
the Higgs detection channel.

Of course, $t\bar t$ background would always be present.
The $WW\to t\bar t$ scattering via  $t$-channel  $b$ quark exchange is
suppressed in phase space compared to
$WW\to h,\; H$ production followed by Higgs decay.
When $h,\; H \to t\bar t$ threshold opens up (not until 400 GeV or so),
one would have genuine $\nu\nu t\bar t \to \nu\nu + bb + 4j$ background.
These again can be distinguished from $\nu\nu tc$ production 
by event topology and jet counting.
Since the $t\bar t/t\bar c$ ratio is not that large \cite{Hou,shalom}
up to $m_{h,\; H} \simeq 500$ GeV, they do not pose a major threat.
However, as seen from Figs. 2 and 3, for Higgs mass beyond
400--450 GeV or so, the signal cross section has also become
too low and the $WW$ background itself may start to become serious.

\subsection{Comparison of Different Processes}

It is of interest to point out  the difference between $\nu\nu tc$ production
and other $t\bar c$ production processes.
The $e^+e^-\to Z^*\to t\bar{c},\; \bar{t}c$\cite{ARS}
process, though rather clean, has very suppressed rate
because the $Z$-$t$-$c$ coupling is loop-induced
(GIM mechanism is intact in the present model context).
It is clear that $e^+e^-\to Z^*\to H(h)Z\to t\bar{c}Z$ has 
identical $\sin^2\alpha$ dependence as the $WW$ fusion process. 
However, this production mechanism is less promising 
since it suffers from $s$-channel suppression (cross 
section decreasing as $1/ s$) at higher energies,
and at the 500 GeV NLC, the rate is already a bit too low \cite{HL}.

The $e^+e^-\to Z^*\to h(H)A$ process is also $s$-channel suppressed,
hence it is not particularly interesting at higher energies.
But it does offer the intriguing signal \cite{HL} of like-sign top quark pairs via 
$h(H)A \to tt\bar c\bar c,\ \bar t\bar t cc$,
signaled by like-sign $W$ plus $b\bar b$ events.
Furthermore, the effects are the largest in this case when
$\sin^2\alpha \to 0$ or $1$, which is complementary 
to the $\sin^2\alpha$ domain of interest, Eq. (13),
for the $e^+e^-\to \nu_e\bar{\nu}_e t\bar{c}$ process.
At the 500--600 GeV NLC, the rates for the two processes
are comparable, both leading to only a handful of clean events.
Thus, though falling short of making a definitive study,
the 500--600 GeV NLC can cover the full range of
$\sin^2\alpha$ and offer us a glimpse of whether
FCNH couplings exist or not.

Turning away from $e^+e^-$ linear colliders,
the process $\mu^+\mu^- \to h,\; H,\; A \to t\bar c$ \cite{ARS2}
at a possible future muon collider capitalizes on 
the larger Higgs-$\mu$-$\mu$ coupling and a sharp Higgs resonance peak.
However, because of the narrow width of the Higgs boson, 
this would demand \cite{HL} precise tunings of the muon energies 
to find the Higgs resonance.
In contrast, the beauty of the $WW$ fusion process of Ref. \cite{shalom}
and discussed here is that no energy scan is necessary.
It is not yet clear whether a high energy muon collider 
can be built or not \cite{mumu}.

Finally, let us compare with prospects at the LHC.
The challenge for $VV\to h,\; H \to tc$ production search 
is the enormous background.
It has been pointed out, however,  that one might be able to 
{\it directly probe for FCNH coupling strengths} via the 
$cg\to tA\to tt\bar c,\; \bar t\bar tc$ production process 
at the LHC \cite{HLMY}, which does not depend on $\sin\alpha$.
Once again there is the intriguing signature of like-sign top quark pairs.
The event rate is not very high since the raw cross section is
at the 80 fb level \cite{NOTE}, and one still needs to make event selection cuts.
Although promising,
background rejection would certainly still be a major issue,
as is almost always the case for interesting new physics
at hadron colliders.
In contrast to the high rate environment of the LHC, however,
all high $p_T$ events at the NLC would be recorded and scrutinized.
We stress that the search for FCNH effects via $tc$ production is really part of the
Higgs program.
By studying the $VV$ fusion processes alone,
the relative large number of events in $\nu\nu tc$ mode (hundreds of events)
and the concurrent study of Higgs boson properties via the $W^+W^-$ and $ZZ$ 
modes (thousands and hundreds of events, respectively)
should allow one to measure the $h,\; H \to tc$ branching ratios,
which in turn can lead to a determination of the FCNH coupling.
Thus, this has the advantage of being a complete program,
and would be complementary to the $cg\to tA$ process at the LHC.
However, since it would only be fruitful for $\sqrt{s} > 1$ TeV,
the fulfillment of the program would certainly come {\it after}
the studies at the LHC.
  
At any rate, 
we expect the study of $\nu\nu tc$ production via $WW$ fusion 
to be quite feasible.
We urge that a dedicated simulation study of this process for the NLC
be carried out.

\section{Conclusion}

In summary, we have extended the work of Ref. [8] 
on $tc$-production via $e^+e^-\to \nu_e\bar{\nu_e} tc$ at the NLC. 
We elucidate that the particularly promising mass range is 
when both $m_h$ and $m_H$ are of order the weak scale. 
This is quite different from the parameter range discussed by 
the authors of Ref. [8] where one of the Higgs is taken to be as heavy as 1 TeV, 
and consequently the $\sigma_{\nu\nu tc}$ they obtained is smaller than ours.     
With Higgs masses in the range of 200--350 GeV, 
we find that $\sigma_{\nu\nu tc}$ could reach almost 10 fb. 
The $\sin^2\alpha$ dependence is mild for $0.1 < \sin^2\alpha < 0.9$, 
and $\sigma_{\nu\nu tc}$ is greater than 1 fb 
as long as $\sqrt{s}\geq 1 \ {\rm TeV}$. 
Given a significant cross section as such, 
this mode should be searched for carefully at 
future $e^+e^-$ Linear Colliders such as the NLC.

\acknowledgements
This work is supported in part by 
National Science Council of R.O.C. under grant numbers 
NSC 86-2112-M-002-019 and NSC 86-2112-M-009-012.

\newpage

\appendix
\section{The helicity method}

The helicity method is particularly suited 
for numerical manipulations of scattering amplitudes. 
For particles with spin, one constructs explicit representions 
for their helicity wave functions so that the relevant
Feynman amplitudes can be written into numerical forms\cite{ONETOP}. 
Consequently the squaring of 
scattering amplitudes may be performed numerically. 

For fermions, we choose the Weyl basis with
the following represention of $\gamma$-matrices:
\begin{equation}
\gamma^0=\pmatrix{
0 & 1 \cr
1 & 0 \cr
}, \ \ 
\gamma^j=\pmatrix{
0 & -\sigma_j \cr
\sigma_j & 0 \cr
},
\ \
\gamma^5=\gamma_5=\pmatrix{
1 & 0 \cr
0 & -1 \cr
},
\end{equation} 
or collectively
\begin{equation}
\gamma^{\mu}=\pmatrix{
0 & \gamma^{\mu}_{+} \cr
\gamma^{\mu}_{-} & 0 \cr
},
\end{equation}
with $\gamma^{\mu}
_{\pm}=\left(1,\mp \vec{\sigma}\right)$.
The chiral projection operator $P_{\pm}=
(1\pm \gamma_5)/2$ is then given by
\begin{equation}
P_+=\pmatrix{
1 & 0 \cr
0 & 0 \cr
},
\ \
P_-=\pmatrix{
0 & 0 \cr
0 & 1 \cr
},
\end{equation}
where $P_+$ and $P_-$ project onto upper and lower components 
of Dirac four-spinors.

In the Weyl basis, the Dirac spinor $u(\vec{p},\lambda)$ for a fermion 
with momentum $\vec{p}$ and helicity $\lambda$ is given by  
\begin{eqnarray}
u(\vec{p},+)&\equiv& \left (\matrix{
u_+(\lambda=+)\cr
u_-(\lambda=+)\cr
}\right )=\left (\matrix{
\omega_{+}\vert \hat{p}+ \rangle\cr
\omega_{-}\vert \hat{p}+ \rangle\cr
}\right ) \nonumber \\
u(\vec{p},-)&\equiv& \left (\matrix{
u_+(\lambda=-)\cr
u_-(\lambda=-)\cr
}\right )=\left (\matrix{
\omega_{-}\vert \hat{p}- \rangle\cr
\omega_{+}\vert \hat{p}- \rangle\cr
}\right ),
\end{eqnarray} 
where $\omega_{\pm}=\sqrt{E\pm \vert \vec{p}
\vert }$ and 
$\vert \hat{p}\pm \rangle$ denote the two-component eigenvectors of 
the helicity operator $h=\vec{p}\cdot \vec{\sigma}/ \vert \vec{p} \vert$ with
\begin{equation}
\vert \hat{p}+\rangle=\left (\matrix{
\cos{\theta\over 2}\cr
e^{i\phi}\sin{\theta\over 2}\cr
}\right ), \ \ 
\vert \hat{p}-\rangle=\left (\matrix{
-e^{-i\phi}\sin{\theta\over 2}\cr
\cos{\theta\over 2}\cr
}\right ),
\end{equation}
where $\theta$ and $\phi$ are angles specifying
the direction of $\vec{p}$, i.e.
\begin{equation}
\vec{p}=\vert \vec{p} \vert \left(\sin\theta
\cos\phi, \sin\theta\sin\phi, \cos\theta \right).
\end{equation}
Similarly, the spinors of anti-fermions in the Weyl basis are given by
\begin{eqnarray}
v(\vec{p},+)&\equiv& \left (\matrix{
v_+(\lambda=+)\cr
v_-(\lambda=+)\cr
}\right )=\left (\matrix{
\omega_{-}\vert \hat{p}- \rangle\cr
-\omega_{+}\vert \hat{p}- \rangle\cr
}\right ) \nonumber \\
v(\vec{p},-)&\equiv& \left (\matrix{
v_+(\lambda=-)\cr
v_-(\lambda=-)\cr
}\right )=\left (\matrix{
-\omega_{+}\vert \hat{p}+ \rangle\cr
\omega_{-}\vert \hat{p}+ \rangle\cr
}\right ).
\end{eqnarray} 
We note that the helicity wave functions of spin 1 particles 
can be constructed out of the two  building blocks: 
$\vert \hat{p}+ \rangle$ and $\vert \hat{p}- \rangle$.  

For any fermion(anti-fermion) line which contains arbitrary numbers 
of interaction vertices with bosons, the associated amplitude 
must be a linear combination of the structures
\begin{equation}
\bar{w}_1\cdots \gamma^{\rho}\gamma^{\nu}
\gamma^{\mu}P_- w_2,
\end{equation}
and
\begin{equation}
\bar{w}_1\cdots \gamma^{\rho}\gamma^{\nu}
\gamma^{\mu}P_+ w_2,
\end{equation}
where $w_1\equiv \left (\matrix{
w_{1+}\cr
w_{1-}\cr
}\right )$ and $w_2\equiv \left (\matrix{
w_{2+}\cr
w_{2-}\cr
}\right )$ can be either $u$ or $v$. 
Note that, for simplicity, we do not specify 
the momentum and helicity dependence of the spinors.
Using the Weyl representations of Dirac
spinors and $\gamma$-matrices, 
the above two structures are simplified into
\begin{equation}
w^{\dagger}_{1\pm}\cdots \gamma^{\rho}_+\gamma^{\nu}_-
\gamma^{\mu}_+ w_{2-},
\end{equation}
and
\begin{equation}
w^{\dagger}_{1\pm}\cdots \gamma^{\rho}_-\gamma^{\nu}_+
\gamma^{\mu}_- w_{2+},
\end{equation}
where the sign in the subscript of $w^{\dagger}_{1\pm}$ 
depends on the number of $\gamma$-matrices inserted between the spinors. 
Using Eqs. (A4) and (A7), one can express
Eqs. (A10) and (A11) as linear combinations of
\begin{equation} 
\left\langle  \hat{p}_1\pm \vert \cdots \gamma^{\rho}_+\gamma^{\nu}_-
\gamma^{\mu}_+ \vert \hat{p}_2\pm \right\rangle, \end{equation}
and 
\begin{equation} 
\left\langle \hat{p}_1\pm \vert \cdots \gamma^{\rho}_-\gamma^{\nu}_+
\gamma^{\mu}_- \vert \hat{p}_2\pm \right\rangle. \end{equation}
Defining the conjugate spinors as:
\begin{equation}
\tilde{\vert \hat{p}\pm \rangle }\equiv
i\sigma_2 (\vert \hat{p}\pm \rangle )^*,
\end{equation}
where
\begin{eqnarray}
\tilde{\vert \hat{p}+ \rangle }
&=&-\vert \hat{p}- \rangle, \ \
\tilde{\vert \hat{p}- \rangle }
=+\vert \hat{p}+ \rangle, \nonumber \\
\tilde{\langle  \hat{p}+ \vert }
&=&-\langle \hat{p}- \vert, \ \
\tilde{\langle \hat{p}- \vert }
=+\langle \hat{p}+ \vert.
\end{eqnarray}
Then, applying the relation
\begin{equation}
\sigma_2(\gamma_{\pm}^{\alpha} )^T\sigma_2=\gamma_{\mp}^{\alpha},
\end{equation}
we have
\begin{equation} 
\left\langle \hat{p}_1\pm \vert \cdots \gamma^{\rho}_+\gamma^{\nu}_-
\gamma^{\mu}_+ \vert \hat{p}_2\pm \right\rangle
= 
\tilde{\langle \hat{p}_2\pm \vert }  \gamma^{\mu}_-\gamma^{\nu}_+
\gamma^{\rho}_- \cdots \tilde{\vert \hat{p}_1\pm \rangle}, 
\end{equation}
and
\begin{equation} 
\left\langle \hat{p}_1\pm \vert \cdots \gamma^{\rho}_-\gamma^{\nu}_+
\gamma^{\mu}_- \vert \hat{p}_2\pm \right\rangle
= 
\tilde{\langle \hat{p}_2\pm \vert }  \gamma^{\mu}_+\gamma^{\nu}_-
\gamma^{\rho}_+ \cdots \tilde{\vert \hat{p}_1\pm \rangle}.
\end{equation}

We now apply the above formalism to calculate 
$e^+(p_1)e^-(p_2)\to  \bar{\nu}_e(p_3) \nu_e(p_4) t(p_t)\bar{c}(p_c)$. 
First, the amplitude for this process has been written in Eq. (4)  
with its fermionic part denoted as $A\cdot B\cdot C$. 
The explicit forms of $A$, $B$ and $C$ as shown in Eqs. (6) and (8)
can be easily obtained by using Eqs. (A4)--(A18). 
Second, we note that the product $A\cdot B$ involves 
a contraction of Lorentz indices associated with matrices 
$\gamma^{\mu}_+$ and $\gamma_{\mu -}$. 
Such contractions can be evaluated easily via the ``Fierz-like" relation
\begin{equation}
(\gamma^{\mu}_+)_{ij}(\gamma_{\mu-})_{kl}
=(\gamma^{\mu}_-)_{ij}(\gamma_{\mu+})_{kl}
=2\delta_{il}\delta_{kj}
\end{equation}
with $i$ ,$j$, $k$ and $l$ being indices in spinor space. 
Indeed, from (A19), we have
\begin{equation}
\left\langle \hat{p}_3+ \vert \gamma_{\mu -} \vert \hat{p}_1+ \right\rangle \cdot
\left\langle \hat{p}_4- \vert \gamma_+^{\mu} \vert \hat{p}_2- \right\rangle  
=2\left\langle \hat{p}_3+  \vert \hat{p}_2- \right\rangle 
\left\langle \hat{p}_4-  \vert \hat{p}_1+ \right\rangle  
\end{equation}
From Eqs. (4), (6) and (8) and Eq. (A20), 
one now has the full helicity amplitudes $iM(\lambda_t,\lambda_c)$ 
for $e^+(p_1)e^-(p_2)\to  \bar{\nu}_e(p_3) \nu_e(p_4) t(p_t)\bar{c}(p_c)$,
which can be easily incorporated into the numerical program 
ONETOP \cite{ONETOP}.

\section{The interference of  Feynman amplitudes}

In this Appendix, we discuss the interference effects of 
scattering amplitudes arising from different neutral Higgs bosons.
Let us use $i{\cal M}^S$ to denote the amplitudes of
$e^+(p_1)e^-(p_2)\to  \bar{\nu}_e(p_3) \nu_e(p_4) t(p_t)\bar{c}(p_c)$ 
contributions from neutral Higgs bosons $S = H$ and $h$.
That is
\begin{equation}
i{\cal M}^S= i{\cal M}(e^+e^-\to \bar{\nu}_e\nu_e S^*(q))
\times{i\over q^2-m_S^2+im_S\Gamma_S}\times
i{\cal M}(S^*(q)\to t\bar{c}), 
\end{equation}
where $S^*(q)$ denotes the off-shell $S$ with momentum $q$. 
The total cross section $\sigma_{\nu\nu t\bar{c}}\equiv 
\sigma(e^+(p_1)e^-(p_2)\to  \bar{\nu}_e(p_3) \nu_e(p_4) t(p_t)\bar{c}(p_c))$, 
is given by
\begin{eqnarray}
\sigma_{\nu\nu t\bar{c}}&=&
   {(2\pi)^4\over 2s}\int {d^3\vec{p}_3\over (2\pi)^32E_3}{d^3\vec{p}_4\over (2\pi)^32E_4}
   {d^3\vec{p}_t\over (2\pi)^32E_t} {d^3\vec{p}_c\over (2\pi)^32E_c}\nonumber \\
 &\times &\delta^4(p_1 +p_2-p_3-p_4-p_t-p_c)\, 
   \vert i{\cal M}^H+i{\cal M}^h \vert^2.
\end{eqnarray}
One can separate $\sigma_{\nu\nu t\bar{c}}$ 
into diagonal and interference terms, i.e.
\begin{equation}
\sigma_{\nu\nu t\bar{c}}=\sigma^H_{\nu\nu t\bar{c}}+\sigma^h_{\nu\nu t\bar{c}}    
+\sigma^{H-h}_{\nu\nu t\bar{c}}.   
\end{equation}
In the narrow width limit $\Gamma_{H(h)}
\ll m_{H(h)}$, it is well known that
\begin{equation}
\sigma^{H(h)}_{\nu\nu t\bar{c}}=\sigma(e^+e^-\to
\bar{\nu}_e\nu_e H(h))\times \mbox{BR}(H(h)\to t\bar{c}).
\end{equation}
However the interference term
$\sigma^{H-h}_{\nu\nu t\bar{c}}$ is more complicated. 
From Eqs. (B1)--(B3), we obtain
\begin{eqnarray}
\sigma^{H-h}_{\nu\nu t\bar{c}}&=&
  -\cos^2\alpha\, \sin^2\alpha {(2\pi)^4\, \over 2s}
  \int {d^3\vec{p}_3\over (2\pi)^32E_3}{d^3\vec{p}_4\over (2\pi)^32E_4}
  {d^3\vec{q}\over (2\pi)^3}\int (2\pi)^3 dq^2 {1\over 2E_q}\nonumber \\
&\times&\delta^4(p_1 +p_2-p_3-p_4-q)\, 
  \vert i{\cal M}(e^+e^-\to \bar{\nu}_e
  \nu_e {H_{\rm SM}}^*(q)\vert^2 \nonumber \\
&\times &2\, \mbox{Re}\left({1\over q^2-m_H^2+im_H\Gamma_H} \ 
  {1\over q^2-m_h^2-im_h\Gamma_h}\right)\nonumber \\
&\times& \int {d^3\vec{p}_t\over (2\pi)^32E_t} {d^3\vec{p}_c\over (2\pi)^32E_c}
  \delta^4(q-p_t-p_c)\, \vert i{\cal M}({H_{\rm SM}}^*(q)\to t\bar{c})\vert^2,
\end{eqnarray} 
where $E_q\equiv \sqrt{q^2+\vec{q}^2}$, and ${H_{\rm SM}}^*(q)$ has been used to replace 
$H^*(q)$ or $h^*(q)$ since we have factored out
the mixing-angle dependence $\cos^2\alpha\,\sin^2\alpha$. 

In general, $H$ and $h$ are not degenerate. 
Without loss of generality, we may assume
$m_H > m_h$ so that $m_H^2-m_h^2=L\times m_h\Gamma_h$
with $L > 0$. Furthermore let us take 
$x\equiv q^2-m_H^2$. With a little algebra,
the propagator part of Eq. (B5) can be written as
\begin{eqnarray}
& & \int (2\pi)^3dq^2\cdots 2\, \mbox{Re}\left({1\over q^2-m_H^2+im_H\Gamma_H}
  \ {1\over q^2-m_h^2-im_h\Gamma_h}\right)\cdots \nonumber \\
&=& \int dx {1\over 2\pi}\cdots \left({1\over x^2+m_H^2
  \Gamma_H^2}+{1\over (x+Lm_h\Gamma_h)^2+m_h^2\Gamma_h^2}\right) (2\pi)^4\cdots\nonumber \\
&-&\int dx {1\over 2\pi}\cdots {(m_H^2\Gamma_H^2
  -2m_H\Gamma_H m_h\Gamma_h+(L^2+1)m_h^2\Gamma_h^2)\over 
  (x^2+m_H^2\Gamma_H^2)\left((x+Lm_h\Gamma_h)^2+
  m_h^2\Gamma_h^2\right)}(2\pi)^4\cdots. 
\end{eqnarray}
If $H$ and $h$ are precisely degenerate, i.e. $L=0$, 
and $\sin^2\alpha=1/2$ which implies $\Gamma_H=\Gamma_h$, 
the second term on the r.h.s. of Eq. (B6) vanishes, 
while the first term eventually gives rise to 
$-(\sigma^H_{\nu\nu t\bar{c}}+\sigma^h_{\nu\nu t\bar{c}})$ 
which cancells completely the diagonal contributions as expected. 

For $L{\ \lower-1.2pt\vbox{\hbox{\rlap{$>$}\lower5pt\vbox{\hbox{$\sim$}}}}\ } 3 $, 
we note that the $(L^2+1)\, m_h^2\Gamma_h^2$ term in Eq. (B6) 
already dominates over both $m_H^2\Gamma_H^2$ and $m_h^2\Gamma_h^2$, 
provided $\Gamma_h^2$ is not overly suppressed by 
too small a $\sin^2\alpha$. 
Therefore, in Eq. (B6), one may neglect the combination $m_H^2\Gamma_H^2-2m_H\Gamma_H m_h\Gamma_h$ 
with respect to $(L^2+1)m_h^2\Gamma_h^2$.
In this approximation, one can show that the two terms on 
the r.h.s of Eq. (B6) lead to a vanishing interference term
$\sigma^{H-h}_{\nu\nu\bar{t}c}$ in the narrow width limit
of $\Gamma_{H(h)}\to 0$.

To see this, note that in the limit of $\Gamma_{H(h)}\to 0$, 
the dominant contributions to the $x$ integration in Eq. (B6) comes 
from the vicinities of $x=-L\, m_h\Gamma_h$ and $x=0$.
For $x\approx -L\,m_h\Gamma_h$, 
we discard the term $1/(x^2+m_H^2\Gamma_H^2)$ on the r.h.s. of (B6)
while the remaining terms are rearranged as follows:
\begin{eqnarray}
& & \int dx {1\over 2\pi}\cdots {1\over (x+L\,m_h\Gamma_h)^2
      +m_h^2\Gamma_h^2}(2\pi)^4 \cdots \nonumber \\
&-&\int dx {1\over 2\pi}\cdots {(L^2+1)\,m_h^2\Gamma_h^2\over 
  (x^2+m_H^2\Gamma_H^2)\left((x+L\,m_h\Gamma_h)^2+
  m_h^2\Gamma_h^2\right)}(2\pi)^4\cdots \nonumber \\
&=& \int dx \cdots {1\over \pi}{m_h\Gamma_h\over (x+L\,m_h\Gamma_h)^2
     +m_h^2\Gamma_h^2}{1\over 2m_h\Gamma_h}(2\pi)^4\cdots; \nonumber \\
&-&\int dx \cdots{1\over \pi} {(L^2+1)\,m_h^2\Gamma_h^2\cdot m_h\Gamma_h
    \over (x^2+m_H^2\Gamma_H^2)\left((x+L\,m_h\Gamma_h)^2+
  m_h^2\Gamma_h^2\right)}{1\over 2m_h\Gamma_h}(2\pi)^4\cdots. 
\end{eqnarray}
In the limit $\Gamma_h\to 0$, the first term on the r.h.s. of  (B7) 
can be simplified by
\begin{equation}
{1\over \pi}{m_h\Gamma_h\over (x+L\,m_h\Gamma_h)^2+m_h^2\Gamma_h^2}
\to \delta(x+L\,m_h\Gamma_h),
\end{equation}   
while the factor $1/2m_h\Gamma_h$ leads to 
BR$(h\to tc)$ when combined with other terms in Eq. (B5). 
Similarly, the second term on the r.h.s. of  (B7) can be simplified by
\begin{eqnarray}
  {1\over \pi} {(L^2+1)\,m_h^2\Gamma_h^2\cdot m_h\Gamma_h\over 
  (x^2+m_H^2\Gamma_H^2)\left((x+L\,m_h\Gamma_h)^2+
  m_h^2\Gamma_h^2\right)}
    &\ \to \ &   {(L^2+1)\,m_h^2\Gamma_h^2\over 
  (L^2m_h^2\Gamma_h^2+m_H^2\Gamma_H^2)}
\delta(x+L\,m_h\Gamma_h)\nonumber \\
     &\simeq & \delta(x+L\,m_h\Gamma_h),
\end{eqnarray}  
where we have used  $L^2\gg 1$.
Clearly the two terms on the r.h.s. of  (B7) cancel completely,
hence $\sigma^{H-h}_{\nu\nu t\bar{c}}$ receives no contribution from 
$x\approx-L\,m_h\Gamma_h$. 
By similar arguments the integration region $x\approx 0$ also gives 
no contributions to $\sigma^{H-h}_{\nu\nu t\bar{c}}$. 
We therefore conclude that $\sigma^{H-h}_{\nu\nu t\bar c} =0$ 
provided we neglect  the combination 
$m_H^2\Gamma_H^2-2m_H\Gamma_H m_h\Gamma_h$
with respect to $(L^2+1)m_h^2\Gamma_h^2$. 
Keeping the $m_H^2\Gamma_H^2-2m_H\Gamma_H m_h\Gamma_h$
term in Eq. (B6), the resulting
$\sigma^{H-h}_{\nu\nu t\bar{c}}$ is at least $O(1/L^2)$ suppressed 
compared to the total diagonal cross section 
$(\sigma^H_{\nu\nu t\bar{c}}+\sigma^h_{\nu\nu t\bar{c}})$.

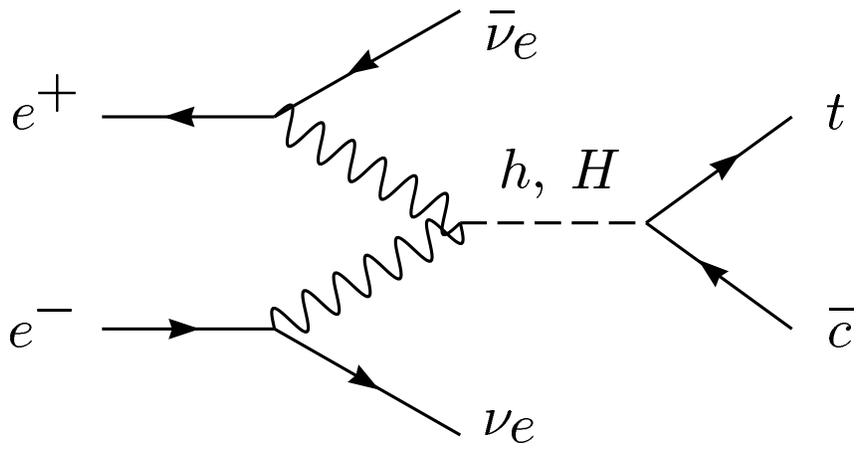
\begin{figure}
\caption{Feynman diagram contributing to the
process $e^+e^-\to \nu_e\bar{\nu}_e t\bar{c}$.}
\label{fig1}
\end{figure}

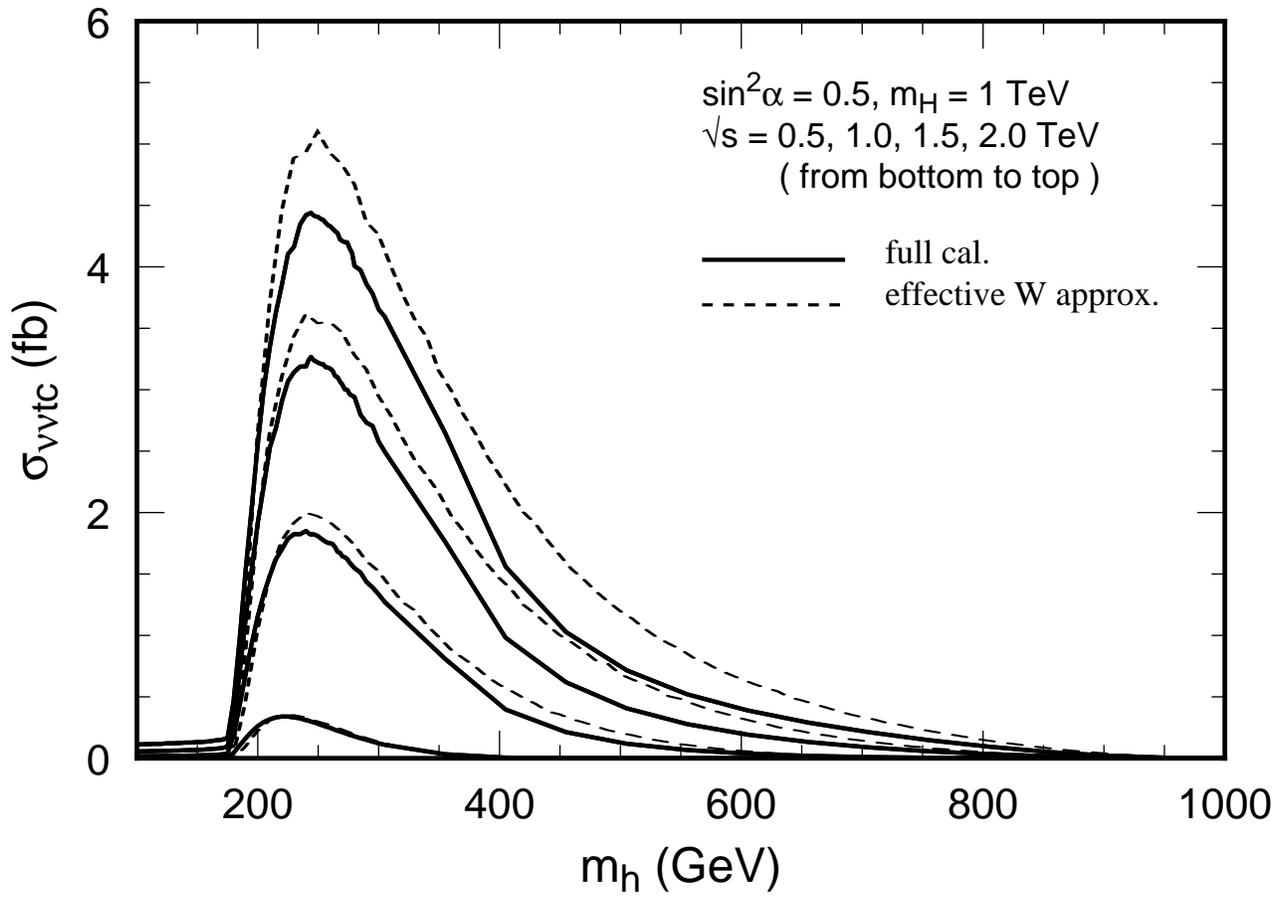
\begin{figure}
\caption{The cross section $\sigma_{\nu\nu tc}$
as a function of $m_h$ with $m_H=1 \ {\rm TeV}$ and $\sin^2\alpha=1/2$
for $s^{1/2} = 0.5,\ 1,\ 1.5,\ 2$ TeV (bottom to top).
Solid lines are for the full calculation, while dashed lines are from
Ref. [8] which uses the effective $W$ approximation.}
\label{fig2}
\end{figure}

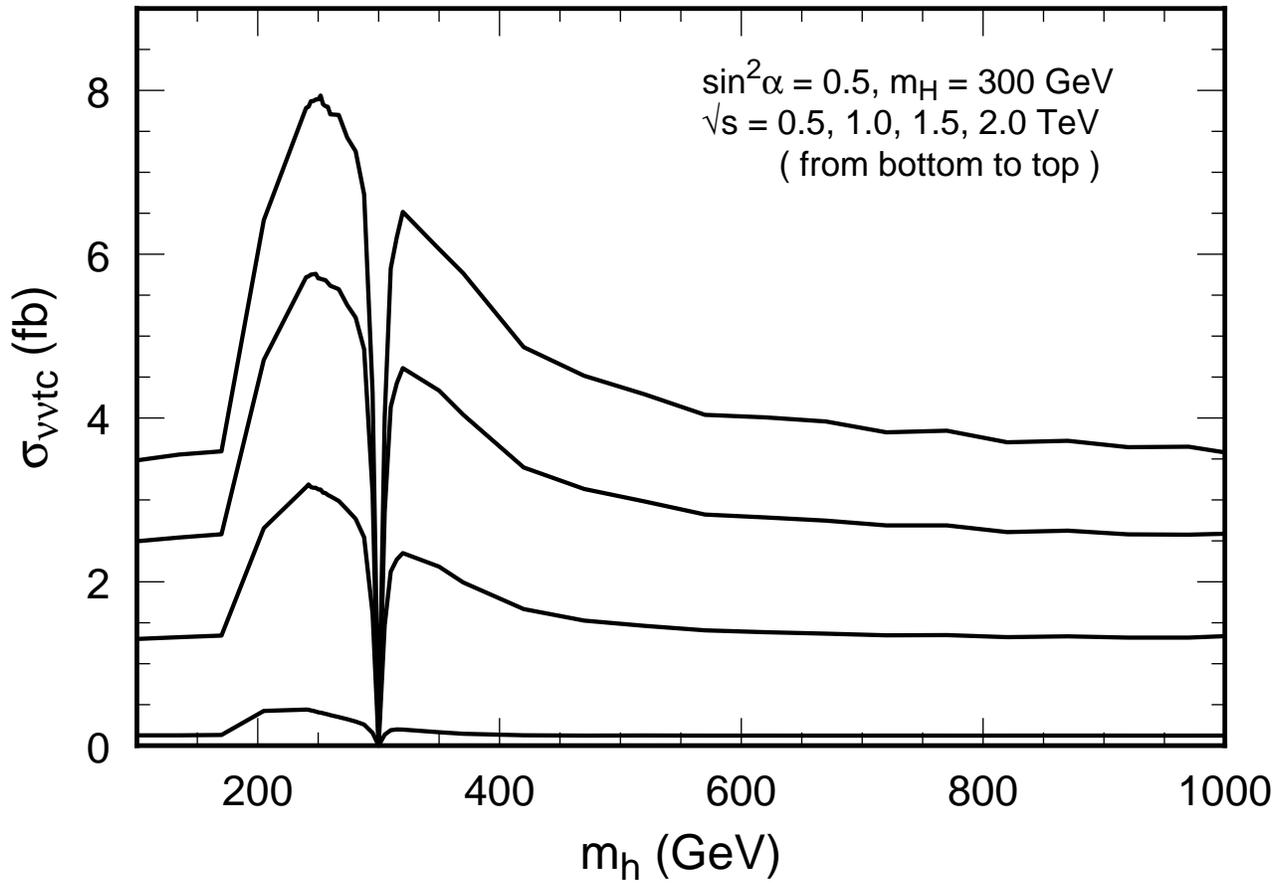
\begin{figure}
\caption{The cross section $\sigma_{\nu\nu tc}$
as a function of $m_h$ with $m_H=300 \ {\rm GeV}$ and $\sin^2\alpha=1/2$
for $s^{1/2} = 0.5,\ 1,\ 1.5,\ 2$ TeV (bottom to top).} 
\label{fig3}
\end{figure}

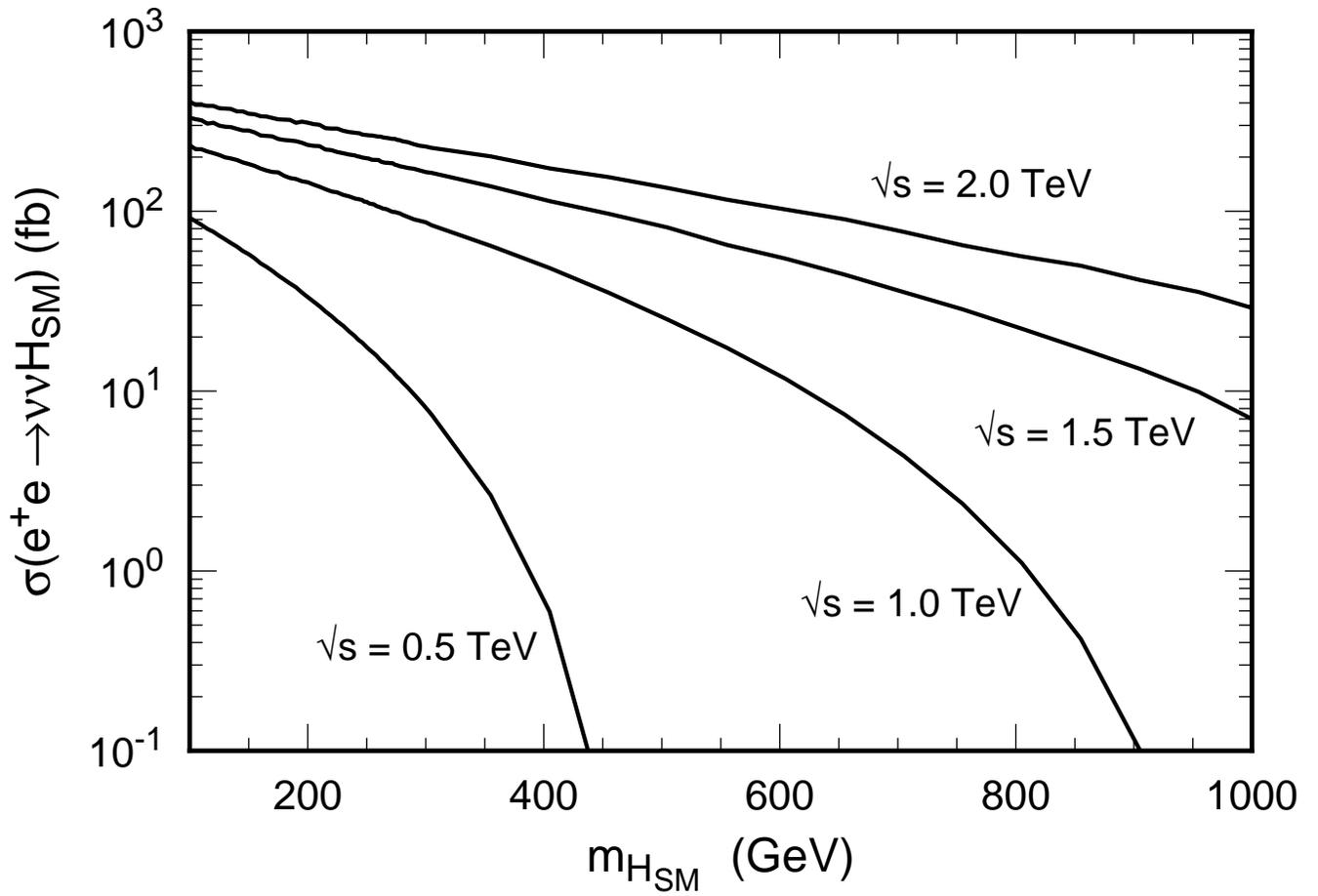
\begin{figure}
\caption{The cross setion $\sigma_{\nu\nu H_{\rm SM}}$ of 
$e^+e^-\to \nu_e\bar{\nu}_e H_{\rm SM}$ as a function of $m_{H_{\rm SM}}$
for $s^{1/2} = 0.5,\ 1,\ 1.5,\ 2$ TeV (bottom to top).}
\label{fig4}
\end{figure}

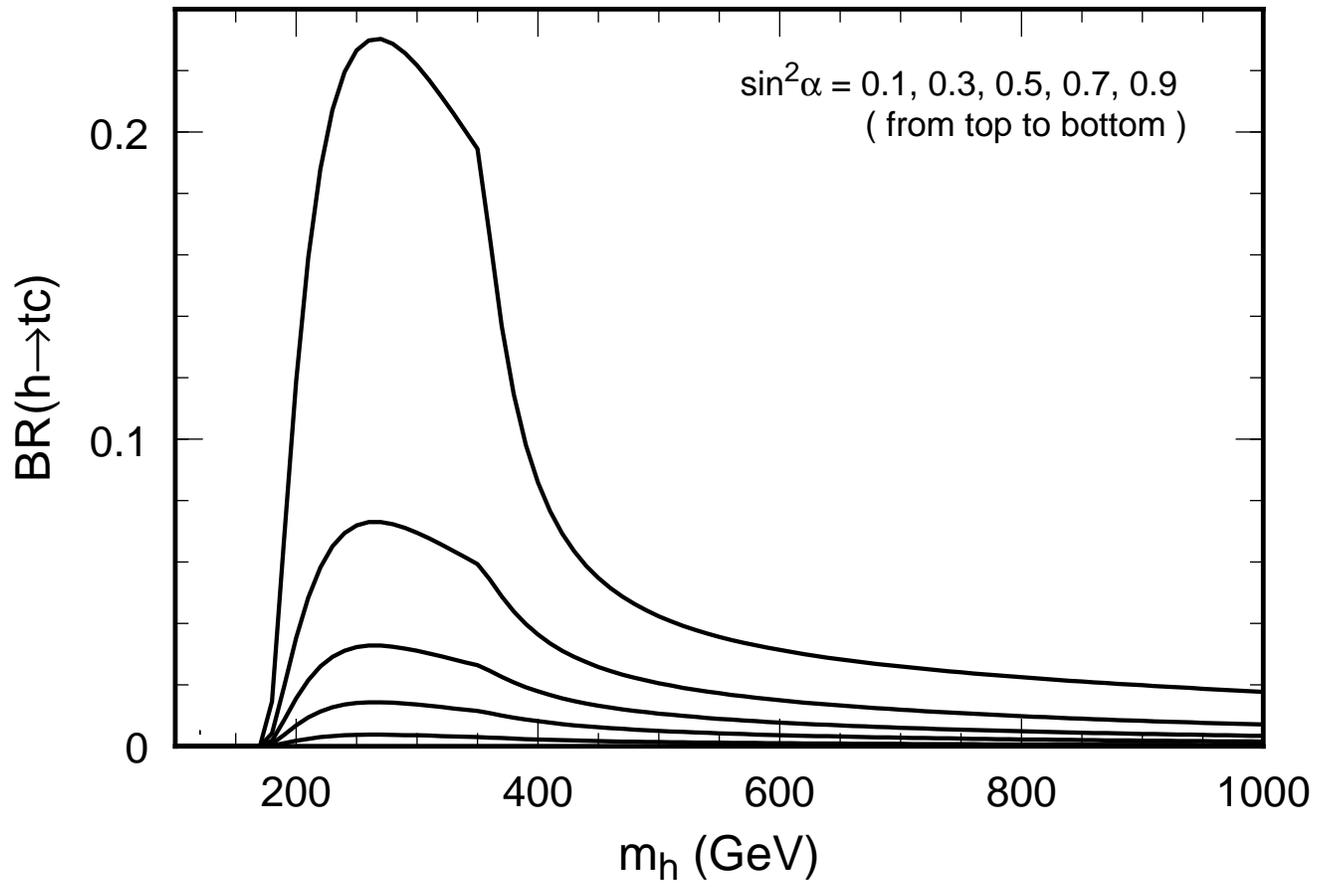
\begin{figure}
\caption{The branching ratio $\mbox{BR}(h\to tc)$ as 
a function of $m_h$ for $\sin^2\alpha = 0.1,\ 0.3,\ 0.5,\ 0.7,\ 0.9$
(top to bottom).}
\label{fig5}
\end{figure}

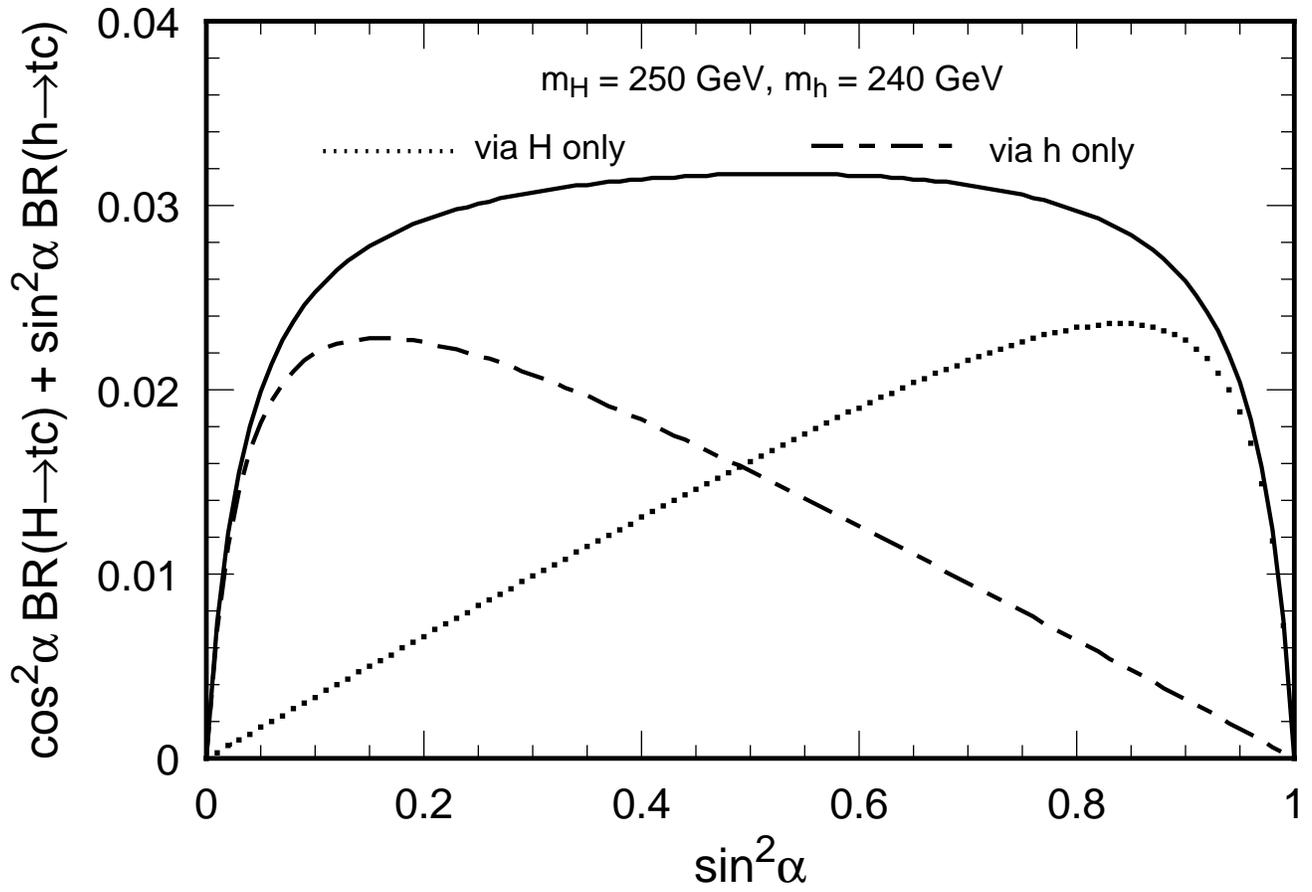
\begin{figure}
\caption{The effective fraction $\cos^2\alpha\, \mbox{BR}(H\to tc)+\sin^2\alpha
\, \mbox{BR}(h\to tc)$ as a function of $\sin^2\alpha$ with $m_H=250 \ {\rm GeV}$ and
$m_h=240 \ {\rm GeV}$.}
\label{fig6}
\end{figure}

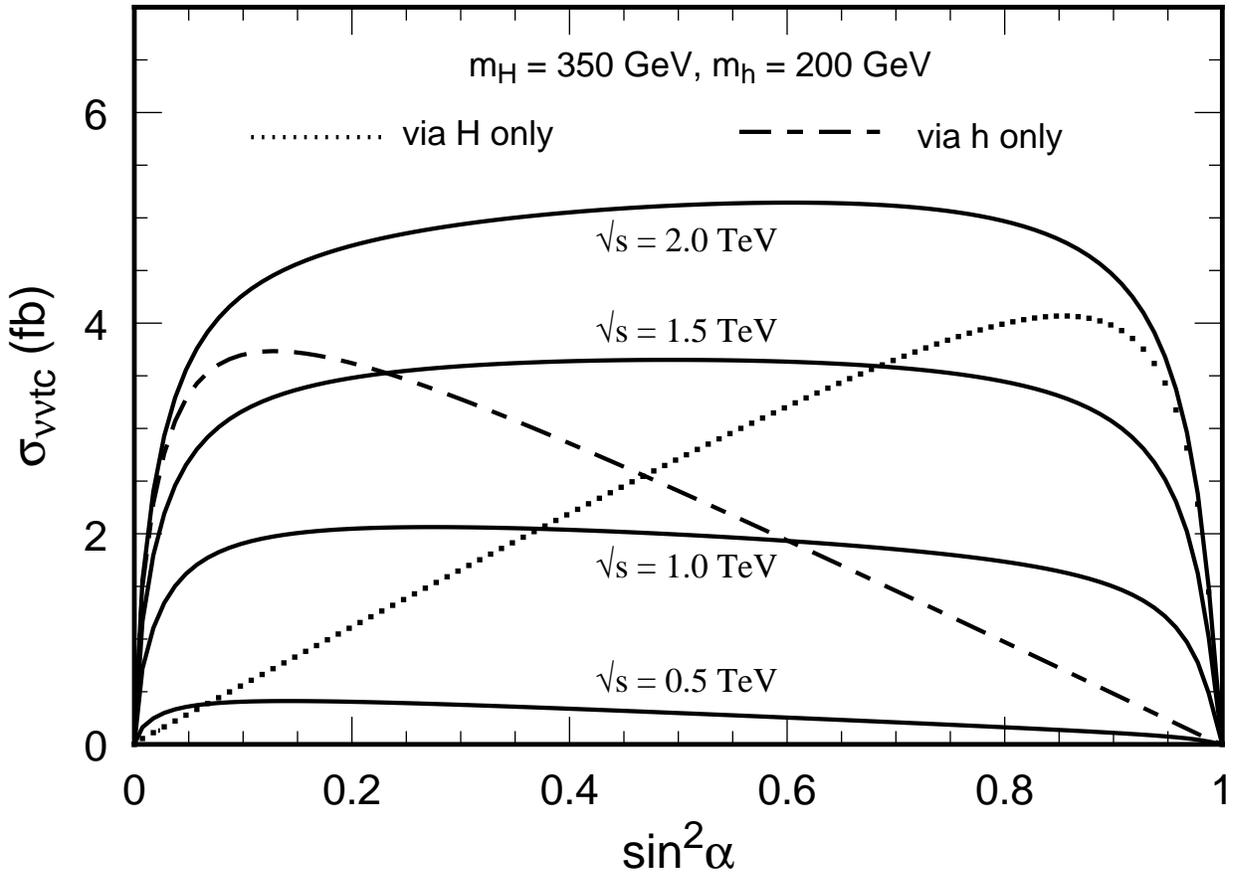
\begin{figure}
\caption{The cross section $\sigma_{\nu\nu tc}$ 
as a function of $\sin^2\alpha$ with $m_H=350
\ {\rm GeV}$ and $m_h=200 \ {\rm GeV}$
for $s^{1/2} = 0.5,\ 1,\ 1.5,\ 2$ TeV (bottom to top).}
\label{fig7}
\end{figure}


\begin{references}
\bibitem{NOTE}
We note that numerical results of Refs. \cite{HL} and \cite{HLMY}
were given in the convention of $f_{tc}=1$, while we have taken
$f_{tc}=\sqrt{2}$ in the current paper for sake of comparison with
Ref.  \cite{shalom}.
\bibitem{NOTE2}
Compare Fig. 2 of Ref.  \cite{Hou},
which studies the case of $h\to t\bar c$ vs. $h\to t\bar t$
in the limit $\sin^2\alpha \to 0$, when $WW$ and $ZZ$ modes are absent.
\end{references}
\end{document}